\documentclass[english]{article}

\usepackage{pifont}

\usepackage{array}
\usepackage{amsmath}
\usepackage{bm}
\usepackage{cancel}
\usepackage{graphicx}

\usepackage{color}
\definecolor{headingcolour}{RGB}{50, 150, 190}

\usepackage{lipsum}
\usepackage{empheq}
\usepackage{babel}

\usepackage[authoryear]{natbib}

\usepackage{setspace}

\usepackage{quattrocento}

\usepackage{sectsty}
\allsectionsfont{\fontfamily{phv}\selectfont}

\usepackage[T1]{fontenc} 

\usepackage[utf8]{inputenc}

\usepackage[margin=1in]{geometry}
\usepackage{fancyhdr}
\usepackage{caption}
\DeclareCaptionFormat{label}{\textbf{#1 #2} #3\hrulefill}
\usepackage[compact]{titlesec}

\usepackage{graphicx}

\usepackage{enumitem}
\setlist{nolistsep}

\usepackage{siunitx}

\renewenvironment{abstract}{%
	\hspace{0.025\linewidth}\begin{minipage}{0.95\textwidth}
		\rule{\textwidth}{1pt}\small\selectfont}
	{\vspace{-0.5em}\par\noindent\rule{\textwidth}{1pt}\end{minipage}\vspace{1em}}
\makeatletter
\renewcommand{\maketitle}{\bgroup\setlength{\parindent}{0pt}
	\thispagestyle{empty}
	\begin{flushleft}
		{\bf \fontfamily{phv}\selectfont \LARGE \@title}
		
		\bf \fontfamily{phv}\selectfont \@author
	\end{flushleft}\egroup
}
\makeatother
\begin{document}
	\title{\fontfamily{phv}\selectfont \Large \textbf{Upscaling transport in heterogeneous media featuring local-scale dispersion: flow channeling, macro-retardation and parameter prediction}}
	\author{\fontfamily{phv}\selectfont Lian Zhou, Scott K. Hansen\footnote{Zuckerberg Institute for Water Research, Ben-Gurion University of the Negev. \texttt{skh@bgu.ac.il}}}
	\date{}
	\doublespacing
	\maketitle
	\onehalfspacing
	\begin{abstract}
		Many theoretical treatments of transport in heterogeneous Darcy flows consider advection only. When local-scale dispersion is neglected, flux-weighting persists over time; mean Lagrangian and Eulerian flow velocity distributions relate simply to each other and to the variance of the underlying hydraulic conductivity field. Local-scale dispersion complicates this relationship, potentially causing initially flux-weighted solute to experience lower-velocity regions as well as Taylor-type macrodispersion due to transverse solute movement between adjacent streamlines. To investigate the interplay of local-scale dispersion with conductivity log-variance, correlation length, and anisotropy, we perform a Monte Carlo study of flow and advective-dispersive transport in spatially-periodic 2D Darcy flows in large-scale, high-resolution  multivariate Gaussian random conductivity fields. We observe flow channeling at all heterogeneity levels and quantify its extent. We find evidence for substantial effective retardation in the upscaled system, associated with increased flow channeling, and observe limited Taylor-type macrodispersion,  which we physically explain. A quasi-constant Lagrangian velocity is achieved within a short distance of release, allowing usage of a simplified continuous-time random walk (CTRW) model we previously proposed in which the transition time distribution is understood as a temporal mapping of unit time in an equivalent system with no flow heterogeneity. The numerical data set is modeled with such a CTRW; we show how dimensionless parameters defining the CTRW transition time distribution are predicted by dimensionless heterogeneity statistics and provide empirical equations for this purpose. 
	\end{abstract}
	\doublespacing
	\section{Introduction}
		When solute moves in a spatially heterogeneous velocity field, particles experience different local transport velocities, resulting in qualitatively different transport behavior from that which is observed under idealized, spatially homogeneous conditions. Because characterization of the flow field is unavoidably coarse, effective upscaled parameters are necessary for predictive modeling. A long-established approach to this is through computation of a macrodispersion coefficient, $D_\infty$, which captures the longitudinal spreading of solute particles in adjacent stream tubes. For small heterogeneity, perturbation approaches lead to an explicit expression for this coefficient \citep{Rubin2003}:
		\begin{equation}
			D_\infty = \left<v_e\right> \sigma^2_{\log K} I_l,
		\end{equation}
		where $\left<v_e\right>$ is the area-averaged Eulerian velocity in the system, and $\sigma^2_{\log K}$ and $I_l$ are respectively the variance and correlation lengths of the underlying multi-Gaussian hydraulic conductivity field. For larger log-variances, numerical studies \citep{Beaudoin2013,Hansen2018} have related $D_\infty$ to non-linear functions of $\sigma^2_{\log K}$. Recently, several papers have adopted a different approach to relation of transport parameters to field statistics. Dispensing with the Eulerian point of view, they have instead proposed a spatial Markov Lagrangian formulation in which velocity transition statistics are derived from the conductivity field. These relationships are used to parameterize random walk particle tracking models in a physically grounded way. Numerous variants exist which assume no local-scale dispersion and flux-weighted injection, invoking the relationship \citep{Dentz2016}:
		\begin{equation}
			p_l(v) = \frac{v p_e(v)}{\left<v_e\right>}.
			\label{eq: eulerian lagrangian pdf}
		\end{equation}
		Here, $p_l(\cdot)$ is the probability density function for the Lagrangian velocity, and $p_e(\cdot)$ is the probability density function for the Eulerian velocity. Using this general approach, \cite{Hakoun2019} related the evolution of Lagrangian velocity PDFs to velocity transition statistics. \cite{Comolli2019} derived effective longitudinal dispersion coefficients, and \cite{DellOca2023} considered transverse macrodispersion in 2D as a result of purely advective fluctuations.
		
		We can see that if flux-weighting is maintained so that \eqref{eq: eulerian lagrangian pdf} is valid then the mean Lagrangian velocity is the same as the (arithmetic) spatial mean Eulerian velocity. Assuming a sufficient travel distance, $\Delta_x$ along a streamline to achieve ergodicity:
		\begin{align}
			\left<v_l\right>^{-1} &= \frac{1}{\Delta_x}\int_0^{\Delta_x} \frac{dx}{v(x)}\\
			&\approx \int \frac{p_l(v_l)}{v_l}\ dv_l \label{eq: fw average}\\
			&= \left<v_e\right>^{-1} \int p_e(v_l)\ dv_l.
		\end{align}
		Thus, $	\left<v_l\right>=\left<v_e\right>$ under these assumptions.
		
		Local-scale dispersion adds complexity in two ways. First, transverse dispersion between adjacent streamlines may cause a secondary, Taylor-type macrodispersion. This was analyzed in detail by \cite{Aquino2021}, who derived a Taylor dispersion coefficient, $D_{T,\infty}$ from analysis of local shear rate, resulting in the scaling relation
		\begin{equation}
			D_{T,\infty} \propto \frac{\bar{v}_e^2 I_t^2}{D}\approx\bar{v}_e \mathrm{Pe}_t I_t,
			\label{eq: taylor D}
		\end{equation} 
		where $\bar{v}_e$ is the \textit{presence-weighted} Eulerian velocity (i.e., the average that is actually experienced by solute at a given point in time), $D$ is a local-scale Fickian transverse dispersion coefficient, $I_t$ is transverse correlation length, and $\mathrm{Pe}_t$ is transverse P\'eclet number. Second, dispersion may lead to net mass transfer of solute from its initial flux-weighted configuration into lower velocity regions. This phenomenon has been shown in a recent numerical and experimental study by \cite{Godoy2019}, who found a substantial additional retardation factor was required to explain upscaled advection in heterogeneous conductivity fields beyond what would be determined by upscaling of local-scale retardation factors, and also by \cite{Cassiraga2005}. By contrast with \eqref{eq: fw average}, if we imagine a periodic system in which concentration is uniform (i.e., particle position is uncorrelated with velocity), the mean Lagrangian velocity over over sufficiently large spatial transitions is the harmonic mean of the Eulerian velocity field, which is smaller than the arithmetic mean, $\left<v_e\right>$.
		
		Dual-porosity systems with mobile-immobile mass transfer (MIMT) are well-studied and exhibit the profound effects of both types of additional phenomena: where a ``flux-weighted'' injection is performed in the mobile phase alone, diffusive interchange with the immobile phase leads to profound effective retardation and Taylor-type dispersion of the solute \citep[e.g.,][]{Neretnieks1980}. Effective dispersion \citep{Michalak2000} and retardation \citep{Uffink2012} coefficients, along with time to effectively Fickian behavior \citep{Hansen2015a} have all been established. Despite this behavior being well-known in the MIMT literature, we find it is not as commonly considered in literature on transport upscaling and macrodispersion for heterogeneous advective systems.
		
		A popular approach to modeling solute transport in heterogeneous media is the continuous-time random walk (CTRW) \citep{Berkowitz2006}, including variants such as the time-domain random walk \citep{Cvetkovic2012, Cvetkovic2014, Hansen2014}. It is desirable to predict the model parameters from the heterogeneity structure. A number of contributions have been made in this area. In the context of a classic CTRW with a truncated power law transition time, \cite{Edery2014} suggested an inverse relationship between power law exponent and $\sigma^2_{\log K}$, and \cite{Nissan2019} reported that it scaled in a power law fashion with longitudinal P\'{e}clet number. In the context of Markovian velocity models, \cite{Meyer2010} derived explicit relationships for drift and dispersion terms, with these ideas being extended by \cite{Dentz2016} and \cite{Comolli2017} and used in a predictive context at the MADE site by \cite{Dentz2022}.
		
		We have previously argued that a simplified CTRW model using average velocity and local-scale dispersion for the advection-like and dispersion-like parameters and unit time constant \citep{Hansen2020, Hansen2020a} can adequately capture MIMT and heterogeneous advection. On this approach, the CTRW transition time distribution, $\psi(t)$, represents a probabilistic  temporal mapping of time taken for a transition with only mean advection and mean local-scale dispersion to the time taken to travel the same distance with all physics present. In \cite{Hansen2022}, we argued that a velocity-invariant $\psi(t)$ is adequate to capture both heterogeneous advection and MIMT, and corroborated these claims against experimental data sets.
		
		In this study, we perform high-resolution particle tracking studies of solute transport with an initially flux-weighted injection in 2D model domains featuring local-scale dispersion as well as Darcian advection in multi-Gaussian hydraulic conductivity fields. We collect statistics on the flow structure and on arrival times of particles at successive planes whose separation exceeds the correlation length, sometimes referred to as \textit{s-Lagrangian} travel times, recording an equal number of transitions for each particle to avoid biased sampling of faster particles. (Throughout this paper, whenever we compute Lagrangian statistics, it will refer to spatially-weighted s-Lagrangian statistics.)  We examine how additional retardation and Taylor dispersion relate to $\sigma^2_{\log K}$, P\'{e}clet number, and anisotropy. Working with non-dimensional equations, we perform automated curve fitting to identify the optimal simplified CTRW model parameters that explain the empirical breakthrough curves.

	\section{Particle tracking simulation}
		We perform flow and particle-tracking simulations in 2D periodic domains. Particles are injected in a flux-weighted fashion and transition times between widely-spaced (relative to correlation length) planes are tabulated for millions of motions, with same number of plane-to-plane transitions are recorded for each particle to avoid bias. 
  
		The 2D model domain for all realizations is 40 m $\times$ 10 m, are discretized on a high-resolution rectangular grid with $2000\times 500$ cells. The grid is equally-divided with a grid spacing of $\delta_x \times \delta_y = $ 2 cm $\times$ 2 cm. By means of the discrete spectral method based on the fast Fourier transform, multiple realizations of base-10 log-normal, random, periodic, hydraulic conductivity field with specified exponential semivariogram model and geometric anisotropy are generated on the same grid. The exponential semivariogram, $C(\Delta_x, \Delta_y)$, is defined:
        \begin{eqnarray}\
             C(\Delta_x, \Delta_y)      &=&\sigma^{2}_{\log K}\exp{\left(-\left[\biggl(\frac{\Delta_x}{I_l}\biggr)^{2}+\biggl(\frac{\Delta_y}{\nu I_l}\biggr)^{2}\right]^\frac{1}{2}\right)}.
        \end{eqnarray}
        Here, $\sigma^{2}_{\log K}$ is the point-wise variance of the log-conductivity field, \(I_l\) its correlation length in the direction of the x-axis quantifying the length at which the field properties are spatially related, and $\nu$ its anisotropy, defining the ratio of the correlation length in the direction of the y-axis to that in the direction of the x-axis. All $\log K$ fields feature a mean value of 0. $\sigma_{\log K}^2$ span values 0.1, 0.3, 0.6, 1, 2, 3, and 5 and feature longitudinal correlation lengths of either 20 cm or 100 cm. For $\sigma^{2}_{\log K} \ge 1$ conductivity field anisotropies of 0.2, 0.4, and 1.0 were considered, with only 0.2 considered for smaller log-variances. For each combination of log-variance, correlation length and anisotropy, a total of four random field realizations were generated.

		\begin{figure}
	    	\centering
	    	\includegraphics[scale=0.6]{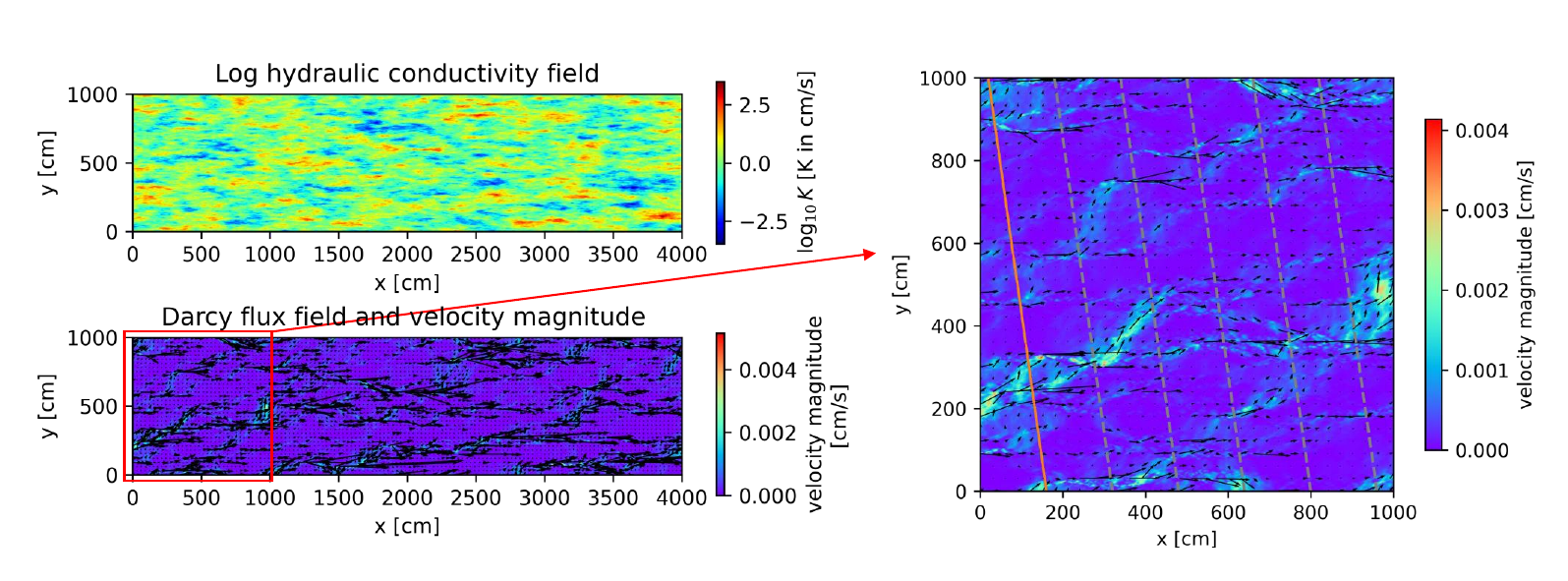}
	    	\caption{Realization of a base-10 log-normal periodic hydraulic conductivity field on 2D domain $\Omega=[0,4000] \times [0,1000]$ with a specified exponential semivariogram model and geometric anisotropy (mean = 0, variance = 0.6, longitudinal correlation length = 100 cm, anisotropy factor = 0.2) and Darcy flux fields solved on the generated field with periodic boundary conditions and a slight enforced mean flow angle $\theta = 8^\circ$ relative to x-axis. Hydraulic conductivity is indicated by the color map. The flux field is indicated by quiver plots whose arrows are interpolated to cell centers, and flux magnitude is indicated by the color map. The dashed lines in the enlarged Darcy flux field plotting show the widely-spaced planes. Particles (shown in orange scatter points) are initially distributed on the injection plane located at the left end of the model domain.}
	    	\label{fig: example}
		\end{figure}
        
        Spatially periodic groundwater flow is simulated using a cell-interface finite volume approach \citep{Zhou2022} with spatially-invariant porosity $\theta=0.25$ assumed throughout. Both the log conductivity field generation and finite-volume groundwater flow solution are generated using freely-available Python package \texttt{periodicgw} \citep{Zhou2021}. This approach allows both the mean Darcy flux vector magnitude and direction for the system to be specified exactly prior to solution. Darcy flux fields are directly solved on the generated heterogeneous conductivity fields, subject to periodic flow boundary conditions and a specified mean Darcy flux vector (including magnitude and direction ). The magnitude of the absolute mean Darcy flux is set to equal to $5.8 \times 10^{-4}$ cm/s, with The direction of the mean Darcy flux is enforced at a slight fixed angle $\theta = 8^{\circ}$ relative to the x-axis (measured counterclockwise from the x-axis) to prevent streamlines from sampling the same heterogeneity on every pass through the domain. For clarity, the mean vector specified is the vector defined by the mean cell-to-cell velocities in the the $x$-direction and the $y$-direction. The magnitude of the mean cell-wise Eulerian velocity is generally slightly larger than that of the specified mean vector, with the difference increasing with heterogeneity. The specified mean is used for computing P\'{e}clet numbers and when performing continuous-time random walk modeling, because it is known a priori and realization-independent. The actual cell-wise mean is used whenever a mean Eulerian velocity is required for normalization when results are graphed. 
        
        At any location, $\mathbf{x}$, within a cell, the advective component of the velocity, $\mathbf{v}(\mathbf{x})$, is computed by linear interpolation of the cell-interface average velocities via the method of \cite{Pollock1988}. Each particle's position is updated in random walk fashion, with deterministic and random components accounting, respectively, for advective and dispersive motion. Particle position is updated according to the Langevin equation
		\begin{equation}
           \mathbf{x}_{j} (t_{j}+\Delta_{t_{j}})  = \mathbf{x}_{j}(t_{j}) + \mathbf{v}(\mathbf{x}_{j})\Delta_{t_j} + \hat{\mathbf{v}}(\mathbf{x}_{j})\xi_l\sqrt{2\alpha_l\left|\mathbf{v^*}(\mathbf{x}_{j})\right|\Delta_{t_j}} + \hat{\mathbf{w}}(\mathbf{x}_{j})\xi_t\sqrt{2\alpha_t\left|\mathbf{v^*}(\mathbf{x}_{j})\right|\Delta_{t_j}}.
		\end{equation}
        Here, $\mathbf{x}_{j}$ is the particle location of the $j$-th particle at time $t_{j}$, $\Delta_{t_j}$ is the current time step duration for the $j$-th particle, and $\mathbf{v}_{j}(\mathbf{x}_{j})$ is the local velocity vector at its location, interpolated using the Pollock method and $\mathbf{v^*}_{j}(\mathbf{x}_{j})$ represents the same local velocity, interpolated using a spatially continuous, bi-linear scheme described below. $\hat{\mathbf{v}}(\mathbf{x}_{j})$ and $\hat{\mathbf{w}}(\mathbf{x}_{j})$ are unit vectors, respectively aligned with, and transverse to, the local flow direction. $\xi_l, \xi_t \sim N(0,1)$ are standard normal random numbers generated anew for each step of each particle. $\Delta_{t_j}$ is dynamically selected for each particle and time step to maintain a small grid Courant number and large grid Péclet number: $\mathrm{Co}_G =  \| \mathbf{v}_{j} \| \Delta_{t_j}/\delta_x < 0.1$ and $\mathrm{Pe}_G = \delta_x\delta_y/2D_L\Delta_{t_j} > 10$. We assume a local longitudinal dispersivity, $\alpha_l$ of 2 cm and a transverse dispersivity, $\alpha_t$ of 0.2 cm for all simulations.
      	
      	A well known challenge in performing particle tracking simulations featuring discretized velocity fields is the potential for artificial transport into lower-velocity regions due to discontinuities in the component of the velocity parallel to a cell interface at said interface \citep{Salamon2006}. Avoiding this is especially pressing for our analyses, as quantifying the effective longitudinal Lagrangian velocity is one of our goals. Thus, we adopt the operator splitting approach advocated and tested by \cite{LaBolle1996}, adapted to a finite-volume code, in which interfacial fluxes are specified. For computation of the local dispersion tensor, we employ a continuous velocity interpolation scheme based on a position-weighted average of fluxes in neighboring cells. In cell $i,j$, define $v^x_{ij}$ as the $x$-directed average velocity (i.e., flux into cell $i,j+1$, divided by $\delta_y \theta$), $f_x$ as the fractional position within a cell in the $x$-direction, and $f_y$ as the fractional position within the cell in the $y$-direction. Then we may compute $\mathbf{v^*}\equiv\left<v^x,v^y\right>$, where
      	\begin{eqnarray}
	      	v^x(f_x,f_y) &= \frac{1-f_x}{2}\left((1-f_y)v^x_{i-1,j-1}+v^x_{i,j-1}+f_y v^x_{i+1,j-1}\right)+\frac{f_x}{2}\left((1-f_y)v^x_{i-1,j}+v^x_{i,j}+f_yv^x_{i+1,j}\right),\\
	      	v^y(f_x,f_y) &= \frac{1-f_y}{2}\left((1-f_x)v^y_{i-1,j-1}+v^y_{i-1,j}+f_x v^y_{i-1,j+1}\right)+\frac{f_y}{2}\left((1-f_x)v^y_{i,j-1}+v^y_{i,j}+f_x v^y_{i,j+1}\right).
      	\end{eqnarray}
      	As dispersion is computed using $\mathbf{v^*}$ only, it varies in a spatially continuous manner, at the cost of being slightly smoothed relative to the interpolated advection, and there is no artificial mass transfer due to discontinuous transverse dispersion across cell interfaces. Also, as the transverse component of an individual particle transition is maintained approximately two orders of magnitude smaller than the scale of velocity variation in the direction transverse to flow by adaptive time stepping and the grid Courant condition, use of the velocity applicable at the start of each step throughout the step is also unlikely to cause significant non-physical mass transfer into low velocity regions.
      	
        We tabulate transition time statistics for particle motion between notional planes separated by orthogonal distance $\Delta_P = 80\ \delta_{x}\cos{\theta}$ and perpendicular to the specified mean Darcy flux direction. In each realization, particles (total number of $M=10^5$) are initially distributed on a injection plane located at the left end of the model domain, following a flux-weighted injection. This means that the number of particles injected at a location is proportional to the local velocity. The coordinates on the y-axis of  cell centers and nodes are extracted from the projection of the injection plane on the y-axis according to previous discretization. The corresponding x-axis coordinates on the plane can be obtained by inverse mapping. From the top to the bottom of model domain, we select $N=1001$ locations on the the injection plane and compute the corresponding local magnitude of Darcy flux $\{ \psi_0, \dots, \psi_{N-1} \}$. The approximately flux-weighted injection is implemented in accordance with:
            \begin{equation}
                \mu_{j} =  \left\lfloor \frac{\psi_{2j-1}}{\sum_{i=1}^{(N-1)/2}\psi_{2i-1}}M \right \rfloor,  j = 1, 2, \dots, \frac{N-1}{2}.
		\end{equation}
        Where the plane are uniformly discretized into $\frac{N-1}{2}$ fragments, and $\mu_{j}$ is the number of particles in the $j$th fragment. Within each fragment, particles are spatially distributed randomly with uniform probability.

        An accumulated displacement vector is maintained for each particle. Once a particle departs from its injection plane or it completes a plane-to=plane transition, this vector's value is set to zero. Its value is updated with each particle motion updated until the projection of corresponding accumulated displacement vector over the vector of mean Darcy flux direction exceeds fixed distance $\Delta_P$, at which point a plane-to-plane transition is recorded. Each particle is simulated until exactly 30 successive plane-to-plane transitions occur, at which time it is removed from the simulation. In this way, we ensure that the Lagrangian travel time statistics are not biased towards faster particles.  
        	
	\section{Simulation results and discussion}
	
	\subsection{Flow channeling and heterogeneity}
\begin{figure}
			\centering
			\includegraphics[width=0.7\linewidth]{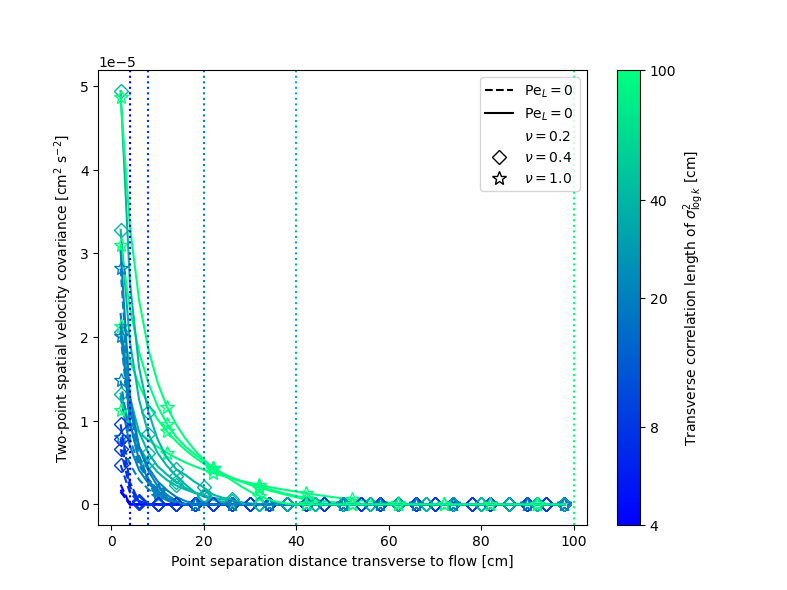}
			\caption{Empirical spatial covariance function of Eulerian velocity in the direction transverse to flow obtained from an ensemble of four realizations for all conductivity log-variances, anisotropies, and correlation lengths simulated. Lines are colored by transverse correlation length of the underlying conductivity fields, and these lengths are also shown, for reference, as appropriately-colored dotted vertical lines.}
			\label{fig: velocity svg}
		\end{figure}

\begin{figure}
			\centering
			\includegraphics[width=0.6\linewidth]{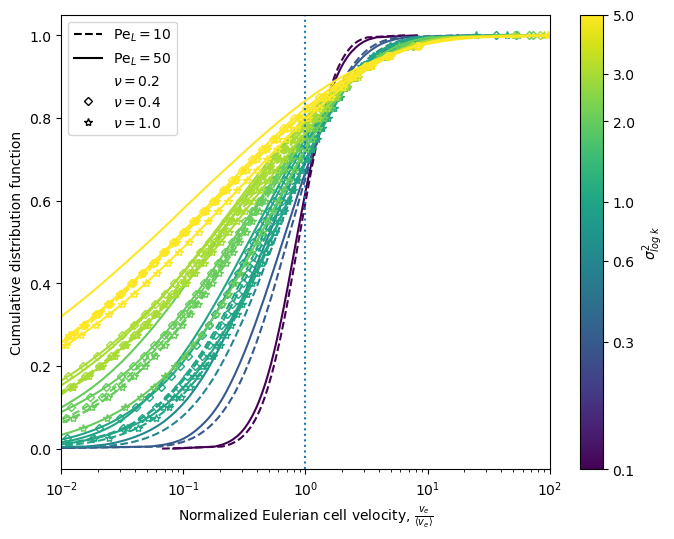}
			\caption{Cumulative distribution of cell-wise velocity magnitude normalized to system mean Eulerian velocity magnitude for various conductivity log-variances, anisotropies, and correlation lengths.}
			\label{fig: velocity CDF}
		\end{figure}

		Previous numerical transport simulations \citep{Edery2014, Hansen2018a} have suggested that in heterogeneous permeability fields, flow develops a strongly channelized pattern, with a comparatively small portion of the subsurface participating in carrying a majority of the groundwater flow. The extent and severity of the channeling will govern how local-scale transverse dispersion influences longitudinal transport. Consequently, we first consider the effect of the heterogeneity meta-parameters, $\sigma_{\log K}^2$, $\nu$, and $I_l$, on flow channeling. Figure \ref{fig: velocity svg} shows empirical velocity spatial covariance functions of the cell-wise Eulerian velocities, averaged from the various realizations with spatial separation in the direction transverse to mean flow. It is apparent that the correlation length of these fields varies with, and is generally significantly smaller than, the corresponding correlation length of the hydraulic conductivity fields. Figure \ref{fig: velocity CDF} shows the cumulative cell-wise velocity magnitude distribution for various combinations of conductivity log-variance, longitudinal correlation length, and correlation length anisotropy. Even at the smallest variance, $\sigma_{\log K}^2=0.1$, significant skewness is apparent, which increases with heterogeneity and inversely with transverse correlation length. When $\sigma_{\log K}^2=5$, a quarter of the domain area features velocities at least two orders of magnitude below the Eulerian mean.
		
	\subsection{Ensemble breakthrough data}
	
		The data from the particle tracking simulations was re-scaled and analyzed in terms of dimensionless groups. We define a dimensionless time, $T = t\left<v_e\right>/\alpha_l$, where $t$ is dimensional time and $\alpha_l$ is the local-scale longitudinal dispersivity. We define $I_l$ as the longitudinal correlation length of the hydraulic conductivity field and $I_t$ as the transverse correlation length. Anisotropy of the hydraulic conductivity field is represented by symbol $\nu \equiv I_t/I_l$. We also define longitudinal and transverse P\'{e}clet numbers, respectively $\mathrm{Pe}_l \equiv I_l/\alpha_l$, and $\mathrm{Pe}_t \equiv \nu \mathrm{Pe}_l$. Throughout the simulations, $\left<v_e\right>$ and $\alpha_l$ are held fixed, so dimensionless times are directly comparable, and the P\'{e}clet numbers are essentially dimensionless proxies for longitudinal and transverse correlation lengths.
		
\begin{figure}
			\centering
			\includegraphics[scale=0.7]{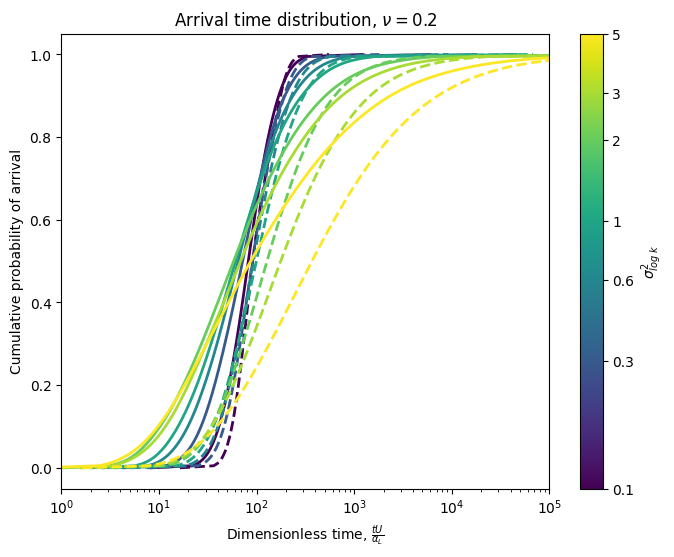}
			\caption{Empirical cumulative breakthrough data for fixed anisotropy, $\nu=0.2$, for a variety of variances (indicated by color) and two values of $\mathrm{Pe}_L$: 50 (solid lines) and 10 (dashed lines).}
			\label{fig:btc fixed aniso}
		\end{figure}

		In Figure \ref{fig:btc fixed aniso}, we see ensemble average cumulative breakthrough curves amalgamated from different realizations corresponding to a fixed $\nu = 0.2$. Cumulative arrival time appears approximately sigmoid when plotted against $\log T$, suggesting mild tailing. In general, the most obvious effect is increased spreading with increasing $\sigma^2_{\log K}$, which is in line with previous literature on macrodispersion. Another obvious effect observed is substantial retardation when $\sigma^2_{\log K} > 1$, for smaller P\'{e}clet numbers. This observation goes against a common assumption that average solute velocity is unaffected by flow heterogeneity and is discussed in detail in the next section. In general arrival time spreading is not profoundly affected by $\mathrm{Pe}_l$, and to the extent an effect is observed, larger $\mathrm{Pe}_l$ is associated with more spreading. This shows that local-scale dispersion is itself not a significant source of spreading. We believe that the additional spreading observed for larger $\mathrm{Pe}_l$ may be spurious: the same $\Delta_x$ between observation planes was employed for both large and small $\mathrm{Pe}_l$, so large $\mathrm{Pe}_l$ simulations manifest less self-averaging over the same distance $\Delta_x$ and feature some amount of ballistic spreading. The effect is small in any event, but worth noting as a possible confounding factor when tabulating Lagrangian statistics.
		
\begin{figure}
			\centering
			\includegraphics[scale=0.7]{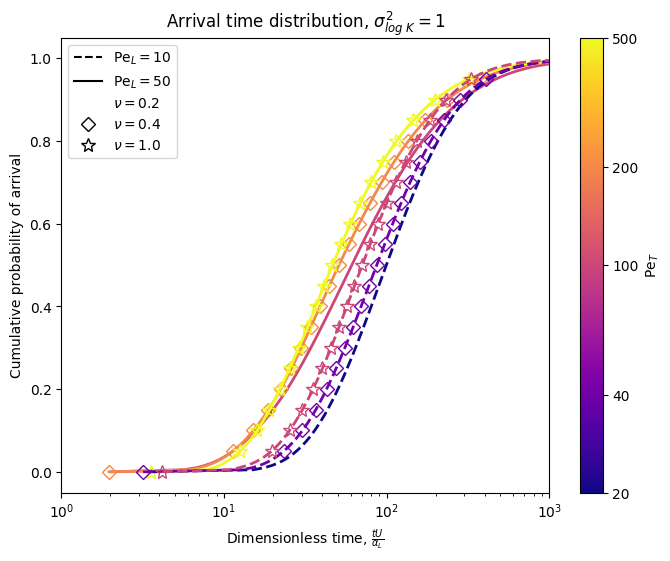}\\
			\includegraphics[scale=0.7]{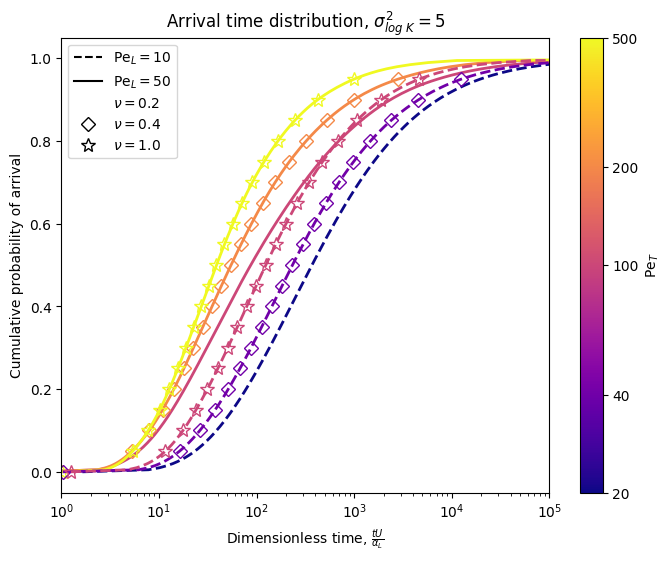}
			\caption{Empirical cumulative breakthrough data for two fixed values of $\sigma^2_{\log K}$.}
			\label{fig:btc fixed variance}
		\end{figure}

\begin{figure}
			\centering
			\includegraphics[width=0.7\linewidth]{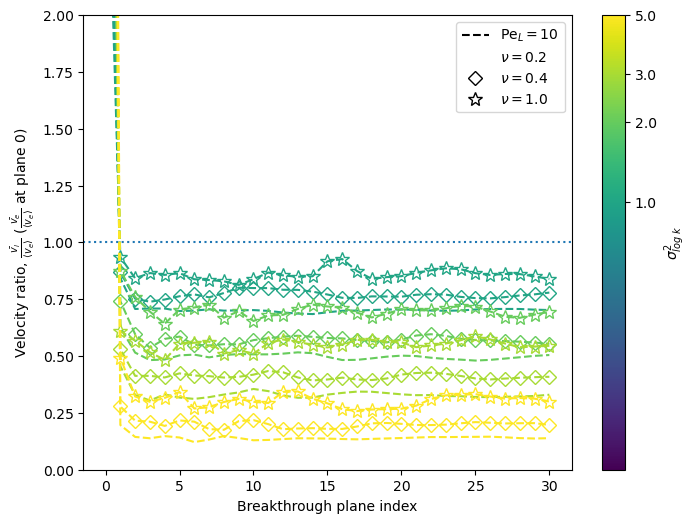}
			\caption{Variation of geometric mean spatial Lagrangian velocity implied by plane-to-plane arrival time, normalized to cell-wise Eulerian mean velocity, versus index of plane of arrival. Value shown for plane 0 (i.e., the injection plane) uses the instantaneous flux-weighted injection velocity in place of the spatial Lagrangian velocity.}
			\label{fig: selected_scenarios}
		\end{figure}

		Figure \ref{fig:btc fixed variance} shows the data in another way, exploring the role of correlation length and anisotropy for two fixed variances. Curves are colored by $\mathrm{Pe}_t$, which clearly illustrates the dominant role of this parameter in determining upscaled retardation. Examining the curves for $\sigma^2_{\log K}=1$, we see that increased correlation length (represented by the P\'eclet number) is associated with greater spreading. This effect is consistent with the presence of Taylor dispersion. At higher heterogeneity, $\sigma^2_{\log K}=5$, this effect is no longer apparent, which requires explanation. Figure \ref{fig: selected_scenarios} shows that the empirical retardation factor, the ratio of the average Lagrangian velocity to the space-averaged Eulerian velocity is as high as 10. We see a clear trend of increasing retardation with increased variance and decreased transverse correlation length. 
		
		
	\subsection{Macro-retardation and Taylor-type macrodispersion}
		Figure \ref{fig: selected_scenarios} shows the ensemble average Lagrangian velocity implied by plane-to-plane transition times normalized to mean Eulerian velocity, as a function of plane index. It is apparent that the Lagrangian velocity, though initially elevated due to flux-weighted injection, quickly drops below the mean Eulerian velocity and becomes quasi-steady within a short distance of injection. Because Lagrangian velocities are well in excess of the harmonic mean velocities for the heterogeneities in question and vary consistently with anisotropy, which does not affect ensemble Eulerian velocity statistics, we conclude the decline in Lagrangian velocity is not due due to diffusive saturation of the immobile region during mobile-immobile mass transfer (MIMT). Rather, we attribute it to the interaction of local-scale dispersion and the heterogeneous advection regime. For a local coordinate system with $x'$ aligned with flow in a flow channel and $y'$ the perpendicular distance from its center line, consider the transverse dispersion term of the advection-dispersion equation:
		\begin{equation}
			\frac{\partial}{\partial y} \left(D_t \frac{\partial c}{\partial y} \right) = \alpha_t \frac{\partial v_{x'}}{\partial y} \frac{\partial c}{\partial y} + \alpha_t v_{x'}\frac{\partial^2 c}{\partial y^2}.
		\end{equation}
		Because both concentration and flow velocity decrease with distance from center line, the first term on the right hand side represents an advective-type flux acting to move material towards the fringes of the flow channels, where longitudinal velocity is slower. We posit that this process, counteracted by periodic re-mixing across the tube during flow focusing and tube merging events, accounts for the observed retardation. This explanation accounts for the greater retardation observed for higher heterogeneity and smaller transverse correlation length conditions, as both act to increase the velocity gradient near the flow channel boundary. We stress this is just a hypothesis: our numerical study recorded only plane arrival times and not a history of particle positions, so within-channel spatial relations could not be directly inspected. Regardless of the explanation, the result that increased channeling is associated with increased retardation was consistent and robust. Owing to the design of our particle tracker, which avoids spurious mass transfer associated with dispersivity discontinuity, we believe it to be a genuine phenomenon, similar to that observed by \cite{Godoy2019}. We may imagine the phenomenon as "macro-retardation", by analogy with macrodispersion: as a retarding phenomenon that is driven by spatial velocity fluctuations, rather than by MIMT.
		
		In general, Taylor-type dispersion increases with increasing P\'{e}clet number, which is apparent in the greater variance associated with the cumulative breakthrough curves in Figure \ref{fig:btc fixed variance} for $\sigma_{\log K}^2=1$. The disappearance of this phenomenon for large variances can be accounted for by the macro-retardation, too. According to \eqref{eq: taylor D}, Taylor-type macrodispersion varies with the square of the Eulerian velocity, weighted by particle presence. By contrast, ordinary macrodispersion scales linearly with Eulerian velocity. Because retardation reduces the presence-weighted Eulerian velocity substantially, $\bar{v}_e^2 \ll \left< v_e \right>$. It is thus reasonable that the effect of Taylor-type macrodispersion becomes negligible with increasing heterogeneity. 
		
\section{CTRW transport model}
		In a recent paper \citep{Hansen2020} we proposed an interpretation of some of the terms of the continuous-time random walk (CTRW) generalized master equation (GME), which allow its 1D form to be written in the following simplified way:
		\begin{equation}
			\frac{\partial c(x,t)}{\partial t} = \int_{0}^{t}M(t-t')\left(-\left<v\right>\frac{\partial c(x,t')}{\partial x}+\alpha\left<v\right>\frac{\partial^2 c(x,t')}{\partial x^2}\right)dt'.
			\label{eq: CTRW GME}
		\end{equation}
		Here, $c\ \mathrm{[ML^{-3}]}$ is concentration, $M(t) \ \mathrm{[T^{-1}]}$ is a temporal memory function, $\left<v\right>\ \mathrm{[LT^{-1}]}$ is mean groundwater velocity, and $\alpha\ \mathrm{[L]}$ is a standard Fickian dispersivity, generated by multiplication of $\bar{v}$ by some fixed, medium-specific dispersivity, $\alpha\ \mathrm{[L]}$; $x\ \mathrm{[L]}$ is spatial coordinate, and $t\ \mathrm{[T]}$ is time. On this approach, $M(t)$ is defined in the Laplace domain according the formula:
		\begin{equation}
			\tilde{M}(s)\equiv\frac{s \tilde{\psi}(s)}{1-\tilde{\psi}(s)},
			\label{eq: CTRW memory}
		\end{equation}
		where superscript tilde denotes the Laplace transform, $s\ \mathrm{[T^{-1}]}$ is the Laplace variable, and $\psi(t) \ \mathrm{[T^{-1}]}$ is the probability distribution function for a subordination mapping representing the total time taken for solute to complete a transition that would have taken unit time under purely advective-dispersive physics as described by $\bar{v}$ and $\alpha$.
		
		The analytic solution of the GME \eqref{eq: CTRW GME} in the Laplace domain for the cumulative arrival of a Dirac solute pulse in a 1D semi-infinite domain has the form \citep{Burnell2017}:
		\begin{equation}
			\tilde{c}(x,s) = \frac{1}{s}\exp\left\{\frac{x}{2\alpha|\bar{v}|}\left[\left<v\right> -\sqrt{\bar{v}^2 +\frac{4\alpha |\bar{v}|s}{\tilde{M}(s)}}\right]\right\}
			\label{eq: analytic solution}
		\end{equation}
		where $\tilde{M}(s)$ is as defined in \eqref{eq: CTRW memory}. For this study, we employ a Gamma-distributed transition time distribution
		\begin{equation}
			\psi(t; k, \theta) = \frac{1}{\Gamma(k)\theta^k} t^{k-1}e^{-t/\theta},
		\end{equation}
		which can be seen to feature a power-law tail with exponential tempering. The two free parameters control the shape, or asymmetry, of the distribution ($k$), and the temporal scaling ($\theta$). This distribution has qualitatively similar features as the popular truncated power law \citep{Berkowitz2006}, and has a simple, well-behaved Laplace transform.
		\begin{equation}
			\hat{\psi}(s;k,\theta)=\left(1+\theta s\right)^{-k}.
			\label{eq: gamma trans}
		\end{equation}
		
		We non-dimensionalize the parameters using the $\alpha_l$ as the characteristic length scale, $\mu_x$ and $\alpha_l/\left<v_e\right>$ as the characteristic timescale, $\mu_t$: these are the natural scales in the periodic domain as hydraulic conductivity heterogeneity goes to zero. Before comparison, we re-scale the empirical breakthrough data in terms of $\mu_t$ and \eqref{eq: analytic solution} in terms of $\mu_x$ and $\mu_t$, to yield
		\begin{equation}
			\tilde{C}(x,s) = \frac{1}{s}\exp\left\{\frac{\Delta_x}{2\alpha_l}\left[1 -\sqrt{1 + 4\left((1+\theta s)^k - 1\right)}\right]\right\}.
			\label{eq: analytic nondim}
		\end{equation}
		The empirical cumulative arrival distributions are discretized by computing the dimensionless times corresponding to 201 quantiles (in half-percent increments) from initial solute arrival to final solute arrival.
		
		We identify parameters $k$ and $\theta$ by automated Levenberg-Marquardt curve fitting of the empirical cumulative arrival time distributions with \eqref{eq: analytic nondim}, using the LMFIT Python package \citep{Newville2024}. We employ the numerical Laplace transform inversion functionality in the Mathematica 14 software \citep{WolframResearch2024} to perform the inversion, invoking it via the Wolfram Client Library for Python. We find this to be the most robust approach for fully-automated calibration due to the sensitivity of numerical inversion accuracy to meta-parameters and need for increased numerical resolution required to invert the transform at early time \citep{Abate2006}. Essentially, for any given value of the parameter vector $\left<k,\theta\right>$, \eqref{eq: analytic nondim} is inverted at the same 201 quantile times, and the integral square (L2) error of $C(x,t)$ as relative to the empirical CDF is computed by Simpson's rule. This error is the quantity that is minimized by the Levenberg-Marquardt algorithm via iterative refinement of the estimated $k$ and $\theta$.
		
\begin{figure}
			\centering
			\includegraphics[scale=0.7]{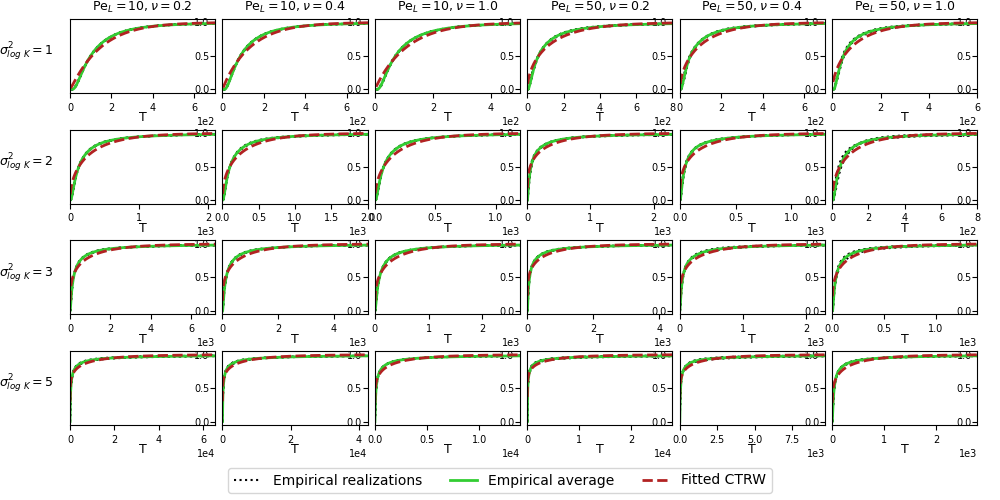}\\
			\caption{Empirical breakthrough curves and optimal CTRW-predicted breakthrough curves with optimized parameters for Gamma-distributed $\psi(t)$. Grid rows represent different heterogeneities, and columns represent different longitudinal correlation lengths and anisotropies. Note that $T$ (dimensionless time) axis scale varies for the various axes in the grid.}
			\label{fig:btc fit grid}
		\end{figure}

		In Figure \ref{fig:btc fit grid}, we show the averaged ensemble cumulative breakthrough curves obtained form the numerical particle tracking simulations overlaid with the best-fit CTRW curve employing \eqref{eq: analytic solution} and \eqref{eq: gamma trans}, as identified with the curve fitting for a variety of log-variances, longitudinal correlation lengths, and anisotropies. We see that the calibrated CTRW model provides qualitatively similar results for all cases, corroborating use of this conceptual approach. Indeed, as the spatially-averaged Eulerian velocity is  employed for the CTRW $v$ and the mean Lagrangian velocity is reduced by a factor of 10 or more (Figure \ref{fig: selected_scenarios}), it is clear that the mapping $\psi$ is doing significant work.
		
\begin{figure}
			\centering
			\includegraphics[scale=0.7]{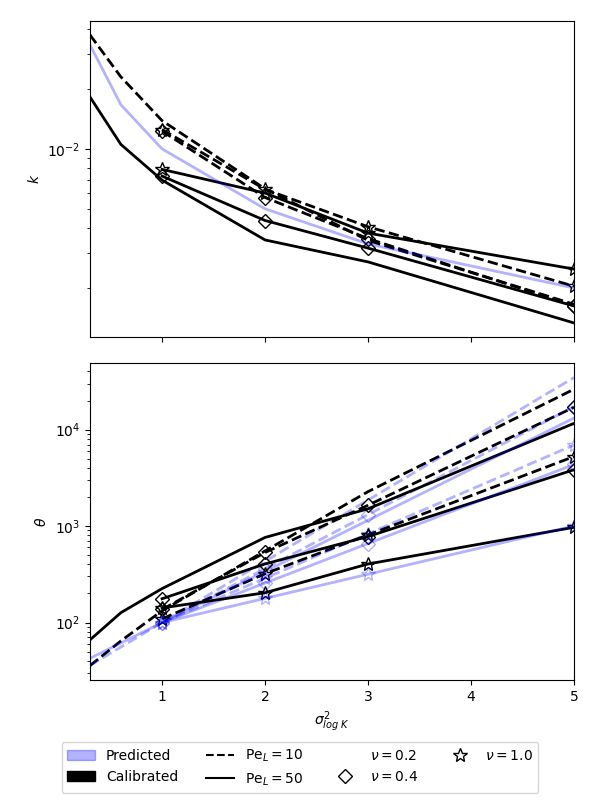}\\
			\caption{Optimized parameters defining $\psi(t;k,\theta)$, as determined by Levenberg-Marquardt curve fitting, shown in black. Blue lines represent predictions encoded in equations (\ref{eq: k pred}-\ref{eq: theta pred}).}
			\label{fig:gamma parameters}
		\end{figure}

		In Figure \ref{fig:gamma parameters}, the fitted $k$ and $\theta$ are shown as functions conductivity field statistics $\sigma^2_{\log K}$, grouped by $\nu$ and $\mathrm{Pe}_l$. Empirical relations for the parameters were fitted, shown in blue in the figure. Defining the convenience variable $\zeta = \sigma^2_{\log K} - 1$, The formulas are:
		\begin{align}
			k &\approx \frac{100}{\sigma_{\log K}^2}, \label{eq: k pred}\\
			\theta &\approx 100\ e^{1.75\zeta}\ \mathrm{Pe}_l^{-.3\zeta}\ \nu^{-(.4 + 3.75\times 10^{-3}(\mathrm{Pe}_l-50))\zeta}.
			\label{eq: theta pred}
		\end{align}
		The difference in complexity of these expressions is because in general a straightforward relationship is seen linking heterogeneity and tailing, for which $k$ is a proxy, and the data set did not support use of a more complex relationship. The macro-retardation, for which time scaling constant $\theta$ is a proxy, has a more complicated relationship involving the inter-relationship of correlation length and anisotropy. We note that the inverse relationship between log-variance and power law exponent $k$ displays the same scaling reported by \cite{Edery2014}. 
			
	
	\section{Summary and conclusions}
	We performed flow and particle tracking simulations in large, high-resolution (40 by 10 m, 2 cm grid cell) 2-D periodic Darcy flow fields, exploring flow channeling and the interaction of local-scale dispersion with log-variance, correlation length, and anisotropy of the hydraulic conductivity field underlying the Darcy flow. We subsequently fit a simplified CTRW transport model to the data and derived empirical relations connecting its parameters to the underlying conductivity field statistics. Key findings: 
	\begin{itemize}
		\item[\ding{227}] Significant flow channeling was consistently observed for even the smallest heterogeneities, which became increasingly profound with decreased transverse correlation length and increased heterogeneity. The transverse correlation lengths of the channels are significantly smaller than those of the corresponding conductivity fields.
		\item[\ding{227}] In the presence of local-scale dispersion, the ultimate mean Lagrangian (i.e., s-Lagrangian) advective velocity reached after a flux-weighted injection typically falls lower than the mean Eulerian velocity, unlike when there is no local-scale dispersion. The degree of retardation increases with degree of flow channeling (i.e., with increased conductivity log-variance and decreased transverse correlation length).
		\item[\ding{227}] Taylor-type macrodispersion was found to be modest, with retardation at higher heterogeneity levels likely limiting its magnitude as it scales with the square of presence-weighted Eulerian velocity.
		\item[\ding{227}] The Lagrangian velocity plateaus at a quasi-steady value quickly, at values well in excess of the harmonic spatial mean of the Eulerian velocity field, implying retardation is driven by heterogeneous advection rather than diffusive saturation of immobile zones. We propose that the process is caused by an effective outward flux towards the low-velocity fringes of the flow channels driven by sharp velocity gradients at those locations, balanced by longitudinal dispersive mixing as the channels widths fluctuate.
		\item[\ding{227}] Because the Lagrangian velocity can be treated as steady, the assumptions of a ``temporal-mapping'' simplified CTRW are satisfied, We calibrate the parameters of a dimensionless CTRW against non-dimensionalized heterogeneity statistics to generate a predictive relation.
	\end{itemize}	
	The success in calibrating the simplified CTRW across a range of statistical structures and the clear trends in the optimal parameters as a function of the underlying statistics provides support for this approach and suggests that the relations derived may be used predicatively.
	
	\section*{Acknowledgments}
	This work was supported by Israel Science Foundation personal research grant 1872/19. LZ was supported by a postdoctoral fellowship from the Blaustein Center for Scientific Cooperation. SKH holds the Helen Ungar Career Development Chair in Desert Hydrogeology.
	
	\section*{Author roles}
	Both authors shared methodology development, investigation, formal analysis, software development, validation, and writing. SKH was additionally responsible for conceptualization and funding acquisition.
	
	\section*{Data availability}
	This work is an entirely computational study that can be replicated from the information given. For convenience, the cumulative plane-to-plane arrival time distributions and associated flow field metadata for each simulation are archived at Zenodo (doi: 10.5281/zenodo.12805291).
	
	\bibliographystyle{plainnat}
	\bibliography{cfs}
	
\end{document}